\documentclass[final,3p,times,square,numbers,sort&compress]{elsarticle}

\usepackage{amssymb}
\usepackage{lipsum}
\usepackage{graphicx}
\usepackage{dcolumn}
\usepackage{bm}
\usepackage{hyperref}
\usepackage{natbib}
\usepackage{color,soul,xcolor}
\usepackage{epsfig}
\usepackage{blindtext}
\usepackage[utf8]{inputenc}
\usepackage{caption,subcaption}
\usepackage{capt-of}
\usepackage{csquotes}
\usepackage{latexsym,amstext,amscd,amsfonts}
\usepackage{setspace}
\usepackage{epic}
\usepackage{mathrsfs, mathtools}
\usepackage{slashed}
\usepackage{enumerate}
\usepackage{fancyhdr}
\usepackage{physics}
\usepackage{dsfont}
\usepackage{xspace} 
\usepackage{float}
\usepackage{multirow}
\usepackage{multicol}
\usepackage{etoolbox}
\AfterEndEnvironment{strip}{\leavevmode}

\newcommand{\be}{\begin{eqnarray}}
\newcommand{\ee}{\end{eqnarray}}

\newcommand{\ba}{\begin{array}}
\newcommand{\ea}{\end{array}}
\newcommand{\beq}{\begin{equation}}
\newcommand{\eeq}{\end{equation}}

\newcommand{\M}{\mathcal{M}}

\journal{Physics Letters B}

\begin{document}

\begin{frontmatter}

\title{Spontaneous CP breaking in $S_3$ flavored Higgs sector with soft-breaking terms}

\author{E. Barradas-Guevara\fnref{label3}}
\ead{barradas@fcfm.buap.mx}
\author{O. F\'elix-Beltr\'an\fnref{label2}}
\ead{olga.felix@correo.buap.mx}
\cortext[cor1]{Corresponding author.}
\author{A. P\'erez Mart\'inez\corref{cor1}\fnref{label2}}
\ead{col424333@colaborador.buap.mx}
\author{E. Rodr\'{\i}guez-J\'auregui\fnref{label4}}%
\ead{ezequiel.rodriguez@correo.fisica.uson.mx}

\author{J. Monta\~no-Peraza\fnref{label4}}%
\ead{javier.montano@unison.mx}

\affiliation[label2]{organization={FCE},
            addressline={Benemerita Universidad Autonoma de Puebla}, 
            city={Puebla},
            postcode={72000}, 
            country={Mexico}}

\affiliation[label3]{organization={FCFM},
            addressline={Benemerita Universidad Autonoma de Puebla}, 
            city={Puebla},
            postcode={72000}, 
            country={Mexico}}

\affiliation[label4]{organization={Depto. Fisica},
            addressline={Universidad de Sonora}, 
            city={Hermosillo, Son. },
            postcode={83000}, 
            country={Mexico}}

\begin{abstract}
We analyze the Higgs sector of the minimal $S_3$-invariant extension of the Standard Model including spontaneous and explicit CP violation arising from the spontaneous electroweak symmetry breaking. This extended Higgs sector includes three SU(2) Higgs doublets with and without complex vev's, which ones providing an interesting scenario to analyze the Higgs masses spectrum, trilinear Higgs self-couplings and CP violation. We present how the spontaneous electroweak symmetry breaking coming from three SU(2) Higgs doublets gives an interesting scenario when spontaneous CP violation arises from the Higgs doublet $\Phi_S$, singlet under $S_3$. Furthermore, a numerical analysis of the Higgs masses and trilinear Higgs self-couplings is presented. Particularly, we find a physical solution for the scenario in which spontaneous CP break is provided by $\Phi_S$. The analysis of soft breaking to $S_3$ symmetry including CP breaking terms opens up a rich parameter space, with Higgs masses and trilinear Higgs self-couplings favoring one of the Higgs bosons as a SM-like type Higgs boson.
\end{abstract}


\end{frontmatter}

\section{Introduction \label{sec:introduction}}

The Higgs boson is a fundamental piece of the Standard Model (SM) providing mass to the gauge bosons and fermions upon spontaneous electroweak symmetry breaking (SSB), and thus preserving the renormalizability of the theory~\cite{Higgs:1964pj,1971NuPhB..35..167T}. In the SM, only one $\textrm{SU(2)}_L$ doublet Higgs field is included, which when acquiring a vacuum expectation value (vev) breaks the $\textrm{SU(2)}_L \otimes \textrm{U(1)}_Y$ symmetry. Although its existence is a fundamental piece of the theory and the SM Higgs potential is very simple and sufficient to describe a realistic model of mass generation, this may not be the final form of the theory. In the SM, each family of fermions enters independently. To understand the replication of generations and to reduce the number of free parameters, more symmetries are usually introduced in the theory. In this direction, interesting work has been done with the addition of discrete symmetries to the SM (see for instance~\cite{Ishimori:2010au,Ishimori:2012zz,Beye:2015wka}  and references therein for a review on the subject).

It is noticeable that many interesting features of the masses and mixing of the SM can be understood using a minimal discrete group, namely the permutation group $S_3$~\cite{Derman:1978iz, Derman:1979nf, Pakvasa:1977in, Pakvasa:1978tx, Mondragon:1998gy, Mondragon:1999jt, Harrison:2003aw, Kubo:2003iw, Kubo:2003pd, Kobayashi:2003fh, Kubo:2004ps, Caravaglios:2005gw, Araki:2005ec, Kubo:2005sr, Koide:2005ep, Grimus:2005mu, Teshima:2005bk, Kimura:2005sx, Koide:2006vs, Mohapatra:2006pu, Kaneko:2007ea,Felix:2006pn,Canales:2013cga,Emmanuel-Costa:2016vej,Vien:2014vka}. 
In the absence of mass, the SM is chiral and invariant with respect to any permutation of the left and right fermionic fields of the same electric charge. Then, to endow with no-null mass terms to all fermions and at the same time preserve the $S_3$ flavor symmetry of the theory, an extended flavored Higgs sector is required with three Higgs $\textrm{SU(2)}$ doublets: one in a singlet and the other two in a doublet of the irreducible representation of $S_3$~\cite{Kubo:2003iw,Mondragon:2007af,Beltran:2009zz}.

Furthermore, the particle observed at the Large Hadron Collider (LHC) corresponds to the SM physical spectrum. It is not known if there is one or many Higgs bosons yet, an indication of the presence of just one Higgs or an extended Higgs sector, as the one proposed in the $S_3$-invariant extension of the Standard Model (SM-$S_3$), could be found in a future running at the LHC~\cite {Barger:2009me,Gupta:2009wn}.
Models with more than one Higgs doublet, with or without supersymmetry, have been studied extensively for a review of supersymmetric and two Higgs-doublet models (THDM)~\cite{Kanemura:2004mg,Djouadi:2005gj,Branco:2011iw}. Different aspects of three and more Higgs doublets models have also been studied, with and without discrete symmetries (see~\cite{Lendvai:1981wn,Adler:1999gv,Ferreira:2008zy,Howl:2009ds}). In particular, in Refs.~\cite{Barroso:2005tq,Barroso:2005da,Barroso:2006pa} it was shown that in two-Higgs doublet models, at tree level, the potential minimum that preserves electric charge and Charge-Parity (CP) symmetries, when it exists, is a stable and global one. Many of these models are not concerned with the unsolved problem of family replication, and thus there is also an analysis of different aspects of the Higgs potential of various discrete flavor groups~\cite{Kubo:2004ps, Hagedorn:2006ir, Tofighi:2009zzb, Morisi:2009sc, Morisi:2010rk,EmmanuelCosta:2007zz,Beltran:2009zz}.
A main theoretical goal is to construct a flavored or extended Higgs potential with SSB in the ground state, which at the same time gives no-null mass term to $W^{\pm},\, Z^{0}$ and fermions of the three observed families.  In this work we consider the symmetry of permutations $S_3$ where the Higgs sector has three Higgs $\textrm{SU(2)}$ doublets fields~\cite{Beltran:2009zz,EmmanuelCosta:2007zz}. The symmetry $S_3$ is the smallest non-Abelian discrete group, which offers a possible explanation of why there are three generations of the quarks and leptons~\cite{Mondragon:1998gy}. Furthermore, the Yukawa couplings of the SM are sufficient to reproduce the masses of the quarks and leptons, and can also make predictions in the neutrino sector~\cite{Kubo:2004ps,Dev:2012ns,Dias:2012bh,Canales:2012dr}.

The discovery of a neutral scalar field with a mass of $125.7\pm 0.4$ GeV~\cite{Aad:2013xqa} in the LHC has been made. With the discovery of the Higgs at CERN, July 4, 2012~\cite{Aad:2012tfa,Chatrchyan:2013lba,Chatrchyan:2012ufa}, our understanding of the physics of particles and fields reach a point at which the SM with one Higgs as a result of the SSB has been confirmed. The next step is setting out the properties of this physical Higgs, mainly its couplings to gauge bosons and fermions, besides its self-couplings~\cite{Miller:2000uz, Baur:2003gp, Dutta:2008bh, Barr:2014sga}. The latest experimental bounds 
$\kappa_h \equiv \lambda_{hhh}/\lambda_{hhh}^{SM}$ reported by ATLAS collaboration~\cite{Cheng:2025aev,ATLAS:2023gzn} ($-1.6<\kappa_h <7.2$ 95\% C.L.) and CMS collaboration~\cite{Hayrapetyan_2025,2025139210} ($-1.4< \kappa_h <7.8$ 95\% C.L.) regarding the couplings of the SM-type Higgs boson provide another way to test the proposed model.

It is important to considered the whole couplings properties in the analysis of extensions of the SM, with Higgs sector that contains more than one $\textrm{SU(2)}$ doublet Higgs field. So it is crucial to experimentally determine if there is only one or there are more scalars (neutral or electrically charged) Higgs states. As we can see, there are still many unsolved answers, of which, the most important are: Why do we observe the generation's replication? Why do we observe a hierarchy of masses between fermions? Where does CP violation come from? As we know, SSB is the mechanism through which the particles acquire mass, but, which is the reason for the large mass difference between the particles of each generation and why there are three generations as well? How can we explain that neutrinos have a small non-vanishing mass? The way we tackled some of  these problems is by considering the permutation symmetry $S_3$, a way to go beyond the SM (BSM)~\cite{Barradas-Guevara:2014yoa, Bhattacharyya:2010hp, Bhattacharyya:2012ze}. 
Extending the Higgs sector with three $\textrm{SU(2)}$ doublet Higgs fields given an invariant potential under permutation symmetry $S_3$, one obtains a greater number of physical states of Higgs bosons~\cite{Barradas-Guevara:2014yoa, Das:2014fea,Gomez-Bock:2021uyu}. Moreover, this permutation symmetry allows us to develop exact and analytical solutions for nine physical Higgs bosons in the normal minimum without CP violation as shown in Refs.~\cite{Barradas-Guevara:2014yoa,Gomez-Bock:2021uyu}. 

As we know, CP violation is one of the distinctive facts of the electroweak interactions, and CP is a possible symmetry of the electroweak Lagrangians, although it has to be broken. Spontaneous CP violation in the scalar sector has been studied in a lot of works prior to extensions of the SM, see~\cite{Lee:1974jb,Chen:2015gaa} and references therein. In particular, extended scalar sectors show spontaneous CP violation given by a relationship between the vacuum expectation values of the Higgs fields. 

The SM-$S_3$ has previously been used to successfully calculate the Higgs masses spectrum and mixings as well as trilinear Higgs self-couplings in the normal minimum ~\cite{EmmanuelCosta:2007zz,Beltran:2009zz}, quark and lepton mixing~\cite{Caravaglios:2005gw,Canales:2013cga}, and flavour changing neutral currents (FCNC)~\cite{Kubo:2003iw,Mondragon:2007af}. The model has three $S_3$ flavoured Higgs fields, $\Phi_{1,2,S}$, which upon acquiring vev's, break the electroweak symmetry. 

SM-$S_3$ has three different stationary points, which can be classified as Normal, Charge Breaking (CB), and Charge Parity Breaking (CPB) minima, according to the vacuum expectation values of the three Higgs fields~\cite{Beltran:2009zz}. An extended Higgs sector opened up the window for CP violation scenarios coming from it. In here, we performed a detailed study of the spontaneous CP breaking conditions of SM-$S_3$, where we have taken only one complex vev assign to the $S_3$ symmetric singlet irreducible representation,  we examine the CP breaking minimization conditions, without explicit breaking of the flavour symmetry, even though it may be spontaneously broken (see Section~\ref{sec:potential}). 

In this work by means of the minimum Higgs potential invariant under $S_3$, we give an analysis focusing on the CP violation, in particular, through  the trilinear Higgs self-couplings of the neutral scalar Higgs bosons spectrum. When the spontaneous symmetry breaking occurs, there is at least a phase after the spontaneous symmetry breaking in some Higgs doublet. The Higgs potential has the characteristic that it is invariant under CP transformations~\cite{Barradas-Guevara:2014yoa}.  Despite having an invariant potential under CP, we can still observe CP violation under certain conditions, which will be analyzed in this work.We can even include in the CP invariant potential terms of soft-breaking to the discrete symmetry $S_3$ that include explicit CP breaking terms.

This paper is organized as follows: in Section~\ref{sec:potential}, the general potential and the different CPB scenarios of the model are shown; in Section~\ref{sec:minandmass}, we showed the minimum conditions from the two different scenarios in which we obtain the trilinear Higgs self-couplings; Section~\ref{Sec:NumAnalisis} contains the numerical computation of the mass states and the trilinear Higgs self-couplings $\lambda_{h_{i}h_{j}h_{k}},\, (i,j,k=2,3,4,5,6)$ where it allowed us to find a Higgs boson as the right like-SM Higgs candidate in each scenario, and finally in Section~\ref{sec:conclusions} we summarize the conclusions of this work.

\section{A general scalar Higgs potential\label{sec:potential}} 
\noindent The Lagrangian ${\cal L}_{\Phi_i}$ of the extended Higgs sector SM-$S_3$ includes three complex $\textrm{SU(2)}$ doublet fields: 
\be
{\cal L}_{\Phi_i} =\left[ D_\mu \Phi_S\right]^2+\left[D_\mu \Phi_1\right]^2
+\left[D_\mu \Phi_2\right]^2-V\left( \Phi_1,\Phi_2,\Phi_S \right),
\label{eq:one}
\ee
where $D_\mu$ is the usual covariant derivative,
$ D_\mu= \partial_\mu-\frac{i}{2}g_2{\tau_a}  {W_{\mu}^a}-\frac{i}{2}g_1B_\mu$, with $g_1$ and $g_2$ standing for the coupling constants $\textrm{U(1)}$ and $\textrm{SU(2)}$ respectively.
The most general Higgs potential $V\left( \Phi_1,\Phi_2,\Phi_S \right)$
invariant under $\textrm{SU(3)}_C \otimes \textrm{SU(2)}_L \otimes \textrm{U(1)}_Y\otimes S_3$ can be written as~\cite{EmmanuelCosta:2007zz,Beltran:2009zz}:

%
\be 
\ba{rcl}
V\left( \Phi_1,\Phi_2,\Phi_S \right)&=&\mu_1^2\left(\Phi^\dagger_1 \Phi_1+ \Phi^\dagger_2 \Phi_2\right)+ \mu_0^2\left(\Phi^\dagger_S \Phi_S\right) +
\frac{a}{2}\left(\Phi^\dagger_S \Phi_S\right)^2 + b\left( \Phi^\dagger_S \Phi_S \right)\left( \Phi^\dagger_1 \Phi_1+\Phi^\dagger_2 \Phi_2 \right) \\
&&+
 \frac{c}{2}\left( \Phi^\dagger_1 \Phi_1+\Phi^\dagger_2 \Phi_2 \right)^2
+
\frac{d}{2}\left(\Phi^\dagger_1 \Phi_2-\Phi^\dagger_2 \Phi_1\right)^2 + e \textit{f}_{ijk}\left(\left(\Phi^\dagger_S \Phi_i\right)\left(\Phi^\dagger_j \Phi_k\right)+\textrm{H.c.}\right)
\\
 &&+
 f\left\{\left(\Phi^\dagger_S \Phi_1\right)\left(\Phi^\dagger_1 \Phi_S\right)+\left(\Phi^\dagger_S \Phi_2\right)
\left(\Phi^\dagger_2 \Phi_S\right) \right\}  +\frac{g}{2}\left\{ \left(\Phi^\dagger_1 \Phi_1-\Phi^\dagger_2 \Phi_2\right) ^2+
\left(\Phi^\dagger_1 \Phi_2+\Phi^\dagger_2 \Phi_1\right)^2\right\}
\\
&&+\frac{h}{2}\bigg\{\left(\Phi^\dagger_S \Phi_1\right)
\left(\Phi^\dagger_S \Phi_1\right)+\left(\Phi^\dagger_S \Phi_2\right)\left(\Phi^\dagger_S
\Phi_2\right) +\left(\Phi^\dagger_1 \Phi_S\right)\left(\Phi^\dagger_1 \Phi_S\right) + \left(\Phi^\dagger_2 \Phi_S\right)\left(\Phi^\dagger_2 \Phi_S\right)\bigg\},
\ea
\label{eq:potential}
\ee

%
\noindent where $f_{112}=f_{121}=f_{211}=-f_{222}=1$, and $\mu_0^2$, $\mu_1^2$ are mass parameters; $a$, $b$, $\dots$, $h$ are real and dimensionless parameters.  We can write down the $\textrm{SU(2)}$ Higgs doublets to include the discrete flavor symmetry $S_3$  as 
\be
\ba{rcl} \label{eq:doubletshiggs}
\Phi_1&=&\frac{1}{\sqrt{2}} \left( \ba{c}
\phi_1+i\phi_4\cr
\phi_7+i\phi_{10}
\ea
\right),\,
\Phi_2=\frac{1}{\sqrt{2}}\left( \ba{c}
\phi_2+i\phi_5\cr
\phi_8+i\phi_{11}\ea
\right), \, 
 \Phi_S=\frac{1}{\sqrt{2}}\left( \ba{c}
\phi_3+i\phi_6\cr
\phi_9+i\phi_{12}\ea
\right).
\ea
\ee
In the analysis, it is better to introduce nine real quadratic forms $x_i$ invariant under $\textrm{SU(2)}\otimes \textrm{U(1)}$ given as
\be\label{eq:xveccomp}
\ba{ccc}
x_1=\Phi^\dagger_1 \Phi_1,&
x_4= {\cal R}\left(\Phi^\dagger_1 \Phi_2\right),&
x_7= {\cal I}\left(\Phi^\dagger_1 \Phi_2\right),\cr
x_2=\Phi^\dagger_2 \Phi_2,&
x_5= {\cal R}\left(\Phi^\dagger_1 \Phi_S\right),&
x_8= {\cal I}\left(\Phi^\dagger_1 \Phi_S\right),\cr
x_3=\Phi^\dagger_S \Phi_S ,&
x_6= {\cal R}\left(\Phi^\dagger_2 \Phi_S\right),&
x_9={\cal I}\left(\Phi^\dagger_2 \Phi_S\right) .
\ea
\ee
Now, it is a simple matter to write down the SM-$S_3$ potential~\eqref{eq:potential},

%
    \be \label{eq:potential1}
\ba{lll}
V(x_1,\dots,x_9)&=&\mu^2_1\left(x_1+x_2 \right)+\mu^2_0 x_3+\frac{a}{2}x^2_3+b\left(x_1+x_2 \right)x_3+\frac{c}{2}\left(x_1+x_2 \right)^2-2dx_7^2+2e \left[ \left(x_1 - x_2\right)x_6+2x_4 x_5\right]\\[.3cm]
&&+f\left(x_5^2+x_6^2+x_8^2+x_9^2 \right)+\frac{g}{2}\left[\left(x_1-x_2 \right)^2+4x_4^2 \right]+h\left(x_5^2+x_6^2-x_8^2-x_9^2 \right),\\
\ea
\ee
%
\noindent and we can rewrite the potential $V(x_1,\dots,x_9)$ and express it in quadratic form as a simple matrix form as
\be 
V({\bf X} )={\bf A}^T{\bf X}+\frac{1}{2}{\bf X}^T{\bf B}{\bf X}.
\label{eq:potential2}
\ee
The vector ${\bf X}$ given by 
\be
{\bf X}^T=\left( x_1, x_2 ,x_3, \dots,x_9 \right),\label{appppa} 
\ee
${\bf A}$ is a mass parameter vector
\begin{eqnarray}\label{eq:amatrix}
 {\bf A}^T=\left(\mu^2_1,\mu^2_1,\mu^2_0,0,0,0,0,0,0 \right)  \label{appppb}
\end{eqnarray}
and ${\bf B}$  is a $9\times9$ real parameter symmetric matrix
\be \label{eq:bmatrix}
{\bf B}= \left( \ba{ccccccccc}
(c+g) & (c-g) &b&0&0&2e&0&0&0\cr
(c-g) & (c+g) & b&0&0&-2e&0&0&0\cr
b&b&a&0&0&0&0&0&0\cr
0&0&0&4g&4e&0&0&0&0\cr
0&0&0&4e&2(f+h)&0&0&0&0\cr
2e&-2e&0&0&0&2(f+h)&0&0&0\cr
0&0&0&0&0&0&-4d&0&0\cr
0&0&0&0&0&0&0&2(f-h)&0\cr
0&0&0&0&0&0&0&0&2(f-h)
\ea
\right).
\ee
The matrix ${\bf B}$ must be positive definite~\cite{VALIAHO198619,Unwin:2011rn}.

The vector $A$ can be called the soft interaction vector, since it is the vector that accompanies the quadratic Higgs interactions, the potential expressed above is invariant under the $S_3$ symmetry, any term added to the vector $A$ would be a term that breaks the $S_3$ symmetry.

The Higgs potential shown in Eq.~\eqref{eq:potential} is an invariant potential under CP, however, we can break the CP invariance if we include complex vev's, $\langle \phi_i \rangle \, \epsilon \, \mathbb{C}$ and CP violation soft-breaking terms, that is, the generalized vector $\mathbf{A}$ can be expressed as
\begin{equation}
\mathbf{A}^{T}=\left( A _{0},A_1,A_2,A_3,A_4,A_5,A_6,A_7,A_8\right).
\end{equation}
For the analysis, we define three scenarios: 
Scenario I, CP invariant; Scenario II, Spontaneous CP violation given by $\langle \phi_i\rangle=v_i e^{i\gamma_i}$ (i=1,2,3), in particular $ \gamma_3 \neq 0$,  $\gamma_1 = \gamma_2 = 0$; and Scenario III,  explicit CP violation through with quadratic soft-breaking terms.

\subsection{Scenario I: CP invariant\label{subsec:CPI}}

For the CP-invariant case, we have several references in the literature \cite{Das:2014fea, Barradas-Guevara:2014yoa, Gomez-Bock:2021uyu}. In these studies, the normal minimum has been taken and the analytical expressions for the masses of the model's bosons have been found. In \cite{Das:2014fea},
 expressions for the stability of the potential are presented, and the existence of a residual symmetry $\mathcal{Z}_{2}$ that appears after the spontaneous breaking of symmetry is mentioned. This residual symmetry shields the appearance of certain couplings since some of the model's bosons are odd under this symmetry. In addition, this work presents the expressions for the unitarity conditions of the 3-Higgs doublet model under the $S_3$ symmetry.

On the other hand, in the reference \cite{Gomez-Bock:2021uyu}, expressions for trilinear couplings are presented, where the existence of the residual symmetry $\mathcal{Z}_2$ is completely evident. We will perform a numerical analysis in the parameter space of the model, taking into account the conditions of unitarity and stability, as well as constraints on the values of the masses of charged bosons related to experimental bounds. This will allow us to obtain the numerical values of the trilinear couplings and give us references to compare with the scenarios we will consider later of spontaneous CP violation and explicit CP violation taking the soft-breaking of the $S_3$ symmetry.

\subsection{Scenario II: Spontaneous CP violation\label{subsec:CPII}}
In the CP conserving case, the vacuum expectation values of the Higgs doublets are taken as real values. This case was carried out in Ref.~\cite{Barradas-Guevara:2014yoa}, which was considered as the normal minimum, where we have adopted for convenience vev's $v_i$ ($i=1,2,3$), 
with $v_i \in \mathcal{R}$. The CP breaking minimum (CPB) we have%
\be
\langle \Phi_i \rangle = \displaystyle\frac{1}{\sqrt{2}} \left( \begin{array}{c} 0 \\ v_i e^{ i \gamma_i} \end{array}\right) \, \qquad
i = 1, 2, 3 \, ,
\ee
where $\gamma_i \in \mathcal{R}$. In general, we can rewrite {\it vev's} in spherical coordinates, where we are going to have two angles related to the real part and the phases in each case. 
\begin{subequations}
\begin{align}
v_1 &= v \cos\phi \sin \theta (\cos(\gamma_1) + i\sin(\gamma_1)),\\
v_2 &= v \sin\phi \sin \theta (\cos(\gamma_2) + i\sin(\gamma_2)), \\
v_3 &= v \cos \theta  (\cos(\gamma_3) + i\sin(\gamma_3)),
\end{align}
\end{subequations}
where $\gamma_i \in \mathcal{R}$. In general, it should satisfy the constraint
$
v= \left(v_1^2 + v_2^2 + v_3^2\right)^{1/2}
$.
The analysis in which we are going to focus is  taking the values following: $ \gamma_3 \neq 0$, and $\gamma_1 = \gamma_2 = 0$. We assume the Higgs vev's are free parameters subject to the constraint.  Furthermore, we are considering the simplest case, assuming that in general we can have more phases that can be absorbed by making a rotation as mentioned in different models of 2HDM~\cite{Gunion:2005ja} and 3HDM~\cite{Emmanuel-Costa:2016vej, Kuncinas:2023ycz}, the goal  of this analysis is to have a CP source that is not removed and to determine how it affects the neutral Higgs self-couplings that appear in the model when one of them is identifying as SM-like Higgs.

The potential parameters in Eq.~\eqref{eq:potential}, specifically the mass parameters $\mu_0^2$ and $\mu_1^2$, may be written in terms of the vev's.
In this scenario, we expect residual symmetry $\mathcal{Z}_2$ to remain present, since including a phase in any of the vevs does not represent an explicit change in the $S_3$ symmetry. For this scenario, since there is spontaneous CP violation when including it in one of the vev's, we will have a mixture of scalar and pseudoscalar bosons. However, we expect to be able to identify two of these bosons as trilinear couplings equal to zero, where the residual symmetry that appears after the spontaneous symmetry breaking becomes evident.

\subsection{Scenario III: Soft-breaking with explicit CP violation\label{subsec:CPIII}}
For Higgs potential invariant under the 
$S_3$ symmetry, we known that $A_0=A_1=\mu _{1}^{2},\,A_2=\mu _{0}^{2}$ and $A_i=0 \,(i\neq 0,1)$.
All these terms lead to the normal minimum, which is characterized by not spontaneously violating CP.
J. Kubo et al.~\cite{Kubo:2004ps} proposed to include soft-breaking terms, which if one only take into account the equivalent term $A_{4}$ as non-zero, one can arrive at a different minimum to normal one. All of this is valid as long as the spontaneous CP violation comes from the Higgs singlet.

In a few words, we will take a single Kubo soft violation parameter which allows us to observe spontaneous CP violation. At the same time, we can support this under the argument that nature, despite not being contained under the $S_3$ symmetry, is still a symmetry close to nature, and the breaking we make to the symmetry in the potential is minimal. We will take the vector $A$ with soft-breaking in the following form,
\be
\ba{rcl}
\mathbf{A}^{T}&=&\left( A_1,A_1,A_0,A_2,A_3,A_4,A_5,A_6,A_7\right),
\ea
\ee
where $A_2,A_3,A_4,A_5,A_6,A_7$ are the soft-breaking terms.

The terms to be included into the potential in the quadratic part that generate the soft break in the potential are as follows: $$\mu^2_{12} \Phi^\dag_1 \Phi_2  + \mu^{*2}_ {12} \Phi^\dag_2 \Phi_1, \, \mu^2_{1s} \Phi^\dag_1 \Phi_s  + \mu^{*2}_ {1s} \Phi^\dag_s \Phi_1, \,  \mu^2_{2s} \Phi^\dag_2 \Phi_s  + \mu^{*2}_ {2s} \Phi^\dag_s \Phi_2,$$ where $\mu^2_{12} =  \tilde{\mu}^2_{12}e^{i\varphi_{12}}$, $\mu^2_{1s} =  \tilde{\mu}^2_{1s}e^{i\varphi_{1s}}$ and $\mu^2_{2s} =  \tilde{\mu}^2_{2s}e^{i\varphi_{2s}}$,  making the expansion in order to identify the extra elements of vector $\mathbf{X}^{T}$ we have: $\tilde{\mu}^2_{12}\cos{\varphi_{12}} x_4$, $\tilde{\mu}^2_{1s}\cos {\varphi_{1s}} x_5$, $\tilde{\mu}^2_{2s}\cos {\varphi_{2s}} x_6$, $\tilde{\mu}^2_{12}\sin{\varphi_{12}} x_7 $, $\tilde{\mu}^2_{1s}\sin{\varphi_{1s}} x_8 $ and  $\tilde{\mu}^2_{2s}\sin{\varphi_{2s}} x_9$. 
Then, we have
\be
\mathbf{A}^{T}&=&\left( \mu _{1}^{2},\mu _{1}^{2},\mu _{0}^{2},\tilde{\mu}^2_{12}\cos{\varphi_{12}},\tilde{\mu}^2_{1s}\cos{\varphi_{1s}},\tilde{\mu}^2_{2s}\cos{\varphi_{2s}},\tilde{\mu}^2_{12}\sin{\varphi_{12}},\tilde{\mu}^2_{1s}\sin{\varphi_{1s}},\tilde{\mu}^2_{2s}\sin{\varphi_{2s}}\right)
\label{eq:vectorAadriana}
\ee
In this scenario we are going to consider the normal minimum with 
\be
\phi_7=v_1,\ \phi_8=v_2,\ \phi_9=v_3,\
\phi_i=0, \,\,\,\, i\neq 7,8,9\, , \nonumber
\ee
where we have adopted for convenience vev's $v_i$ ($i=1,2,3$), with $v_i \in \mathcal{R}$.
In general, we can rewrite {\it vev's} in spherical coordinates, where we are going to have two angles,  
\begin{subequations}
\begin{align}
v_1 &= v \cos\phi \sin \theta ,\\
v_2 &= v \sin\phi \sin \theta , \\
v_3 &= v \cos \theta,
\end{align}
\end{subequations}
%

\section{Minimum conditions and mass matrices}\label{sec:minandmass}

\subsection{Scenario II: Minimum conditions ($\gamma_3 \neq 0$ and $\gamma_1=\gamma_2=0$) \label{subsec:scen2}}
In this section, we present the minimum conditions and the parameter space analysis for the considered scenario. The minimization conditions give us six equations  determined by demanding of $\partial V/\partial \phi_i\mid_{{min}}=0$.
We denote $M_i{(\gamma_3)} \equiv \partial V/\partial \phi_i\mid_{min}$.

The minimum conditions for the case when we take $\gamma_3 \neq 0 $ and the other two equal to zero are:
\begin{subequations}
\begin{align}
    M_7({\gamma_3}) &=\frac{1}{2} v_1 \left(v_3^2 \left(C_{\gamma_3}^2 (b+f+h)+S_{\gamma_3}^2 (b+f-h)\right)+(c+g) \left(v_1^2+v_2^2\right)+6 C_{\gamma_3} e v_2 v_3+2 \mu_1^2\right), \label{eq:condmin2}\\M_8({\gamma_3}) &=  \frac{1}{2} \left(2 \mu_1^2 v_2+(c+g)v_2 \left(v_1^2+v_2^2\right)+3 C_{\gamma_3} e \left(v_1^2-v_2^2\right) v_3+\left(C_{\gamma_3}^2 (b+f+h)+S_{\gamma_3}^2(b+f-h) \right) v_2 v_3^2\right) ,\\M_9({\gamma_3}) &=  \frac{1}{2} \left(e \left(3 v_1^2 v_2-v_2^3\right)+C_{\gamma_3} v_3 \left(2 \mu_0^2+(b+f+h) \left(v_1^2+v_2^2\right)+a \left(C_{\gamma_3}^2+S_{\gamma_3}^2\right) v_3^2 \right)\right),\\ 
    M_{10}({\gamma_3}) &=  S_{\gamma_3} v_1 v_3 \left(e v_2+C_{\gamma_3} h v_3\right),\\
    M_{11}({\gamma_3}) &=  \frac{1}{2} S_{\gamma_3} v_3 \left(e \left(v_1^2-v_2^2\right) +2 C_{\gamma_3} h v_2 v_3\right), \\ 
    M_{12}({\gamma_3}) &= \frac{1}{2} S_{\gamma_3} v_3 \left(2 \mu_0^2 +(b+f-h) \left(v_1^2+v_2^2\right)+a \left(C_{\gamma_3}^2+S_{\gamma_3}^2\right) v_3^2\right) ,
    \label{eq:condmin2a}
\end{align}
\end{subequations}
\noindent where $C_{\gamma_3}= \cos\gamma_3$  and $S_{\gamma_3}=\sin \gamma_3$.

Using Eqs.~\eqref{eq:condmin2}-\eqref{eq:condmin2a} we have 
\begin{subequations}
\begin{align}
\mu_1^2  &= -2 (b + f - h) v_2^2 - \frac{1}{2} a \left(C_{\gamma_3}^2 + S_{\gamma_3}^2\right) v_3^2,
   \\
\mu_0^2  &= -2 (c + g) v_2^2 - \frac{1}{2} \left(C_{\gamma_3}^2 (b + f - 5 h) + (b + f - h) S_{\gamma_3}^2\right) v_3^2, \\
e &= -\frac{C_{\gamma_3} h v_3}{v_2}, \\
v_1 &=  \sqrt{3}v_2,\label{eq:solconmind} 
\end{align}
\end{subequations}
 $\mu_0^2$ and $\mu_1^2$ has a dependence on the CPB parameter $\gamma_3$. As we can see in Eq.~\eqref{eq:solconmind} , that condition reveals the residual symmetry $\mathcal{Z}_2$. We showed the results for different scenarios where CP-violation was realized. In next section,  we computed numerically the Higgs mass matrix and Higgs mass eigenvalues for each scenario.

\subsubsection{Higgs masses \label{sec:higgsmasses}}
The Higgs mass matrix is obtained from the computation of the second derivatives of the Higgs potential, 
Eq.~\eqref{eq:potential}. There are twelve real Higgs fields $\phi_i$, and the corresponding Higgs mass matrix is a 12 $\times$ 12 real matrix, then 
\be
(\mathcal{M}^2_H)_{ij} = \displaystyle\left.\displaystyle\frac{1}{2}\displaystyle\frac{\partial^2 V}{\partial\phi_i\partial\phi_j}\right|_{\textrm{CPBmin}} ,
\ee
with $i,j={1,2,\dots,12}$. We have 
\be\label{eq:matrizmasa}
\mathcal{ M}^2_H = \hbox{diag}\left({\bf M}_{C}^2, {\bf M}_{N}^2   \right) \, ,
\ee
with ${\bf M}_{C}^2$ corresponding to the mass matrix of electrically charged Higgs bosons and ${\bf M}_{N}^2$ to the neutral Higgs mass matrix, which are
the $6\times 6$ symmetric and Hermitian sub-matrices. 

In general, we have a matrix for charged and neutral Higgs bosons,  where
\be
{\bf M}^2_{C} = \left(
\begin{array}{cc}
{\bf {M}^2_C}_{11} & {\bf {M}^2_C}_{12}  \\ [.4cm]
{\bf {M}^2_C}_{21} & {\bf {M}^2_C}_{22} 
\end{array} \right), \label{eq:matrizcargada}
\ee
which should satisfy the constraint 
\begin{subequations}
\begin{alignat}{2}
 {\bf {M}^2_C}_{22} &= {\bf {M}^2_C}_{11}, \\
{\bf {M}^2_C}_{21}& = - {\bf {M}^2_C}_{12} .
\end{alignat} 
\label{eq:masascargadas}
\end{subequations}

The neutral Higgs mass matrix is given by
\be
{\bf M}^2_{N} = \left(
\begin{array}{cc}
{\bf {M}^2_N}_{11} & {\bf {M}^2_N}_{12} \\[.4cm] 
{\bf {M}^2_N}_{21} & {\bf {M}^2_N}_{22}
\end{array} \right), \label{eq:matrizneutra}
\ee
with
\begin{subequations}
\begin{alignat}{2}
 {\bf {M}^2_N}_{22}& \neq {\bf {M}^2_N}_{11}, \\[.4cm]
{\bf {M}^2_N}_{21} &= {\bf {M}^{2}_N}^T_{12}. 
\end{alignat} 
\label{eq:masasneutras}
\end{subequations}
Here, ${\bf {M}^{2}_N}^T_{12}$ is the transposed matrix of ${\bf {M}^2_N}_{12}$. 
For the three scenarios, the restrictions~\eqref{eq:masascargadas} and~\eqref{eq:masasneutras} were met. The Higgs masses are obtained by diagonalizing the matrices~\eqref{eq:matrizcargada} and~\eqref{eq:matrizneutra}, for each of the scenarios. How can we know each scenario has got a physically possible situation. In Ref.~\ref{apendiceA}, the expressions of the mass matrices are shown, along with the analytical expressions for the charged Higgs bosons mass eigenstates. For the mass-neutral states we got five Higgs mass eigenvalues that we can not stablish which one correspond to scalar or pseudo-scalar state. It is important to mention that by diagonalizing the $6 \times 6$ mass matrix for neutral Higgs states, the theoretical mass eigenstates are obtained implementing a Mathematica program. However, these are very extensive expressions, therefore we will not show these expressions and will only present their numerical analysis in Section~\ref{Sec:NumAnalisis}.
 Then, we compared that to the Higgs masses and trilinear Higgs self-couplings numerical analysis.

For this model with CP violation arising from the Higgs $S_3$ doublet sector  ($\gamma_3 \neq 0$), among the nine physical Higgs fields, we have four charged bosons which are mass degenerate two by two and four non-degenerated bosons in the neutral sector (see~\ref{apendiceA}. Nevertheless, when CP breaking arises from vev related to the $S_3$ Higgs singlet, we found a physical scenario with three Goldstone bosons, which can give mass to vector bosons $W^{\pm}$ and $Z^0$, with a massless photon and nine physical Higgs fields. At least one neutral Higgs should have a mass of 125.7 $\pm$ 0.4 GeV while the remaining eight additional Higgs states are candidates for new particles. This scenario provides a strong motivation to extend the analysis to CPB phenomenology arising from spontaneous electroweak symmetry breaking. We denote the masses of these Higgs charged bosons as $M_{C_i}$ ($i=1,2$) and $M_{H_j}$ ($j=1,\dots,5$) for the neutral masses. 

\subsection{Scenario III: Minimum conditions $t=\varphi_{12,1s,2s}$ \label{subsec:scen3}}
In this section, we present the minimum conditions and the parameter space analysis for Scenario III. For this, we  consider the simplest case taking just the same phase for the extra terms that were shown in Sec.~\ref{subsec:CPIII} ($t=\varphi_{12},\varphi_{1s},\varphi_{2s}$). In this particular case, the vector $\mathbf{A}^{T}$ is 
{\footnotesize
\begin{align}
\mathbf{A}^{T}=\left( \mu _{1}^{2},\mu _{1}^{2},\mu _{0}^{2},\tilde{\mu}^2_{12}\cos{t},\tilde{\mu}^2_{1s}\cos{t},\tilde{\mu}^2_{2s}\cos{t},\tilde{\mu}^2_{12}\sin{t},\tilde{\mu}^2_{1s}\sin{t},\tilde{\mu}^2_{2s}\sin{t}\right).
\end{align}
}

The minimization conditions give us six equations  determined by demanding of $\partial V/\partial \phi_i\mid_{{min}}=0$.
We denote $M_i{(t)} \equiv \partial V/\partial \phi_i\mid_{min}$. Let's take the particular case,
\begin{subequations}
\begin{align}
    M_7(t) &= \frac{1}{2} v_1 \left(2 \mu^2_1 + (c + g) \left(v_1^2 + v_2^2\right) + 6 e v_2 v_3 + (b + f + h) v_3^2\right) + \left(\tilde{\mu}^2_{12} v_2 + \tilde{\mu}^2_{1s} v_3\right) \cos t,  \label{eq:condmin3}\\
    M_8(t) &=  \frac{1}{2} \left(2 \mu^2_1 v_2 + (c + g) v_2 \left(v_1^2 + v_2^2\right) +  3 e \left(v_1^2 - v_2^2\right)w3 + (b + f + h) v_2 v_3^2 + 
   2 \left(\tilde{\mu}^2_{12} v_1 + \tilde{\mu}^2_{2s} v_3\right) \cos t \right), \\
    M_9(t) &=  \frac{1}{2} \left(-e v_2 \left(-3 v_1^2 + v_2^2\right) + 2 \mu^2_0 v_3 + (b + f + h) \left(v_1^2 + v_2^2\right) v_3 + a v_3^3 + 2 \left(\tilde{\mu}^2_{1s} v_1 + \tilde{\mu}^2_{2s} v_2\right) \cos t \right),\\ 
    M_{10}(t) &=  \left(\tilde{\mu}^2_{12} v_2 + \tilde{\mu}^2_{1s} v_3 \right) \sin t, \\
    M_{11}(t) &=  \left(-\tilde{\mu}^2_{12}v_1 +\tilde{\mu}^2_{2s} v_3 \right) \sin t, \\ 
    M_{12}(t) &=  \left(-\tilde{\mu}^2_{1s} v_1 - \tilde{\mu}^2_{2s} v_2\right) \sin t. \label{eq:condmin3a}
    \end{align}
\end{subequations}
\noindent Using Eqs.~\eqref{eq:condmin3}-\eqref{eq:condmin3a} we have  
\begin{subequations}
\begin{align}
\mu_1^2 &= \frac{1}{2} \left(-(c + g) \left(v_1^2 + v_2^2\right) - 6 e v_2 v_3 - (b + f + h) v_3^2\right), \label{eq:conmua}\\ 
\mu^2_0 &=  -\frac{e \left(3 v_1^2 v_2 - v_2^3\right) + (b + f + h) \left(v_1^2 + v_2^2\right) v_3 + a v_3^3}{2 v_3}, \label{eq:conmub} \\
\tilde{\mu}^2_{12} &=  -\frac{ 3 e \left(v_1^2 -3 v_2^2\right) v_3 \sec t}{4 v_1}, \label{eq:conmuc} \\
\tilde{\mu}^2_{1s} &= \frac{3 e v_2 \left(v_1^2 - 3 v_2^2\right) \sec t }{4 v_1}, \label{eq:conmud} \\ 
\tilde{\mu}^2_{2s} &=  -\frac{3}{4}  e \left(v_1^2 - 3 v_2^2\right) \sec t.  \label{eq:conmue}
\end{align}
\end{subequations}
\noindent From Eqs.~ \eqref{eq:conmuc}-\eqref{eq:conmue} we can found the following relations:
\begin{subequations}
\begin{align}
v_1^2 &=  3v_2^2  - l\frac{\tilde{\mu}^2_{12}}{v_3}, \label{eq:v11}\\
v_1^2 &=  3v_2^2  + l\frac{\tilde{\mu}^2_{1s}}{v_2},\label{eq:v12}\\
v_1^2 &=  3v_2^2  - l\frac{\tilde{\mu}^2_{2s}}{v_1}, \label{eq:v13}
\end{align}
\end{subequations}
where $l= \frac{4v_1 \cos t}{3e}$. Therefore, the soft-breaking of symmetry $S_3$ is reflected in relations  \eqref{eq:v11}- \eqref{eq:v13}, since when we have exact symmetry, we have $v_1^2 = 3 v_2^2$, which is also evidence of the residual symmetry $\mathcal{Z}_2$, but by including the terms that softly break the symmetry, we can see that a term is added to the relation between $v_1$ and $v_2$, which must be small. In reality, we have $v_1^2 = 3v_2^2 + \delta^2$. 

Since we no longer have residual symmetry $\mathcal{Z}_2$, we expect all trilinear couplings to be nonzero.

Related to the Higgs mass matrix, as in the private subsection, we are going to have the $12 \times 12$ real matrix, divided into two diagonal blocks of $6 \times 6$ matrices, the first block associated with charged bosons and the second block associated with neutral states. However, the analytical expressions are not possible to obtain, so the analysis to the mass eigenvalues as the trilinear self-couplings are going to be exclusive numerical.

\section{Numerical analysis}\label{Sec:NumAnalisis}
To perform the numerical analysis, pseudo-random numbers were generated for the potential parameters and the angles of the vev's. Unitarity and stability constraints were imposed~\cite{Das:2014fea}, and values greater than 80 GeV were taken for the charged Higgs masses.

In order to do the numerical analysis we are going to take into spherical coordinates the vev's, {\it i.e.}, $v_1= v\cos \varphi \sin\theta, v_2= \sin \varphi\sin\theta, v_3=v\cos\theta $. Then, for the three scenarios we have two angles, $\varphi$ and $\theta$, and $v= 246$ GeV, fixed by the theory. However, in the Scenarios I and II, where the symmetry $S_3$ is complete, the angle $\varphi$ is fixed due to the residual symmetry $\mathcal{Z}_2$, $\varphi= \frac{\pi}{6}$, in those scenarios we just have the angle $\theta$ as a free parameter. On the other hand, in the Scenario III we have both angles as free parameters, $\theta$ gives us the mix between the vev's related to the $S_3$ doublet and single when the symmetry $S_3$ is complete. Thus, we take $\theta$ as independent parameter. Specifically, $\text{Log}(\tan(\theta))$ ($0<\theta \leqslant \frac{\pi}{2}$). In each scenario, one of the neutral Higgs bosons was restricted to taking mass values in the range of 120 to 130 GeV. Similarly, the values of its rate to the trilinear couplings were restricted to the range of -3 to 8 as the experimental limits have the values~\cite{CMS:2024awa} . This was done to ensure that there are values for the model parameters such that one of the neutral bosons corresponds to the Higgs boson of the SM, and in this way, to be able to see the mass spectrum of the other scalar bosons as well as the expected values for their trilinear couplings. In Figure~\ref{fig:S3masstrili} we show the numerical results of scalar neutral Higgs masses and the corresponding rate of trilinear self-couplings $\kappa_h \equiv \lambda_{hhh}/\lambda_{hhh}^{SM}$ for each defined scenario. 

The range of the values for the different parameters is as follow: $a\in [0,16.5]_I, [2,16]_{II}, [0, 16]_{III},  b \in [-1.8, 16.5]_I,$ $ [-1.5, 9.5]_{II}, [-4, 11.5]_{III},$ $ c \in [0, 14]_I, [1.5, 13.2]_{II}, [1, 11.5]_{III}, d \in [-11, 9.5]_I, $  $[-9, 6.5]_{II}, [-10, 5.5]_{III}, e \in [-6, 0]_I,$ $ [-4.5, 0]_{III}, f \in [-16, 4]_I, [-8.7, 4.2]_{II}, [-11.5, 9]_{III}, g \in [-11, 9.5]_I, $ $[-8.5, 3.4]_{II}, [-9, 6]_{III}, \ h \in [-10, 5]_I, [0.5, 9.5]_{II},$ $ [-5.5, 4]_{III} $, each subindex indicate the values for each scenario.

\subsection{Scenario I: CP invariant\label{sec:normalminimum}}
 Figure~\ref{fig:S3masstrili}(a) shows the neutral Higgs masses $m_{h_0, H_1, H_2}$ of the model. For this case, we have the analytical expressions of the masses and the trilinear self-couplings $\lambda_{hhh}$ in Ref.~\cite{Gomez-Bock:2021uyu}, so we just evaluate the expressions. In particular, $\lambda_{h_0h_0h_0}=0$ for this scenario, since $h_0$ it is odd under the symmetry $\mathcal{Z}_2$, whereas in Figure~\ref{fig:S3masstrili}(b) just the values of $\kappa_{H_1}$ and $\kappa_{H_2}$ appear. As we can see in Figure~\ref{fig:S3masstrili}, the numerical values for $\textrm{Log}(\text{tan} (\theta))$, the restriction for the mass and the trilinear self-couplings of $H_2$ are implemented, are restricted to a range of -0.75 to 1.4, which implies that values for the angle $\theta$ will not take values close to zero, nor will it take values close to $\pi/2$.

In this scenario, the most likely boson to be the SM-like Higgs boson is $H_2$, since according with the mass expressions,  $m_{H_1}\gtrsim m_{H_2}$, for this reason, in this scenario, $m_{H_2}$ has been chosen to restrict its mass values to a range of 120 to 130 GeV. As we can observe in Figure~\ref{fig:S3masstrili}(a), the mass values for the $m_{H_1}$ are above 125 GeV, and  $m_{h_0}$  has values below 125 GeV; however, as mentioned, this boson has zero trilinear coupling.

Figure~\ref{fig:S3masstrili}(b) shows the values $\kappa_{h}$, the Higgs boson $H_2$ has been restricted to taking values between -3 and 8, making it aligns with the current experimental limits, as we can see the values for the trilinear self-couplings of $H_1$ have much more dispersed values compared to the values to which $H_2$ is restricted.

\subsection{Scenario II: spontaneous CP violation}
\begin{figure}[H]
\subfloat[]{{\includegraphics[width=0.48\textwidth]{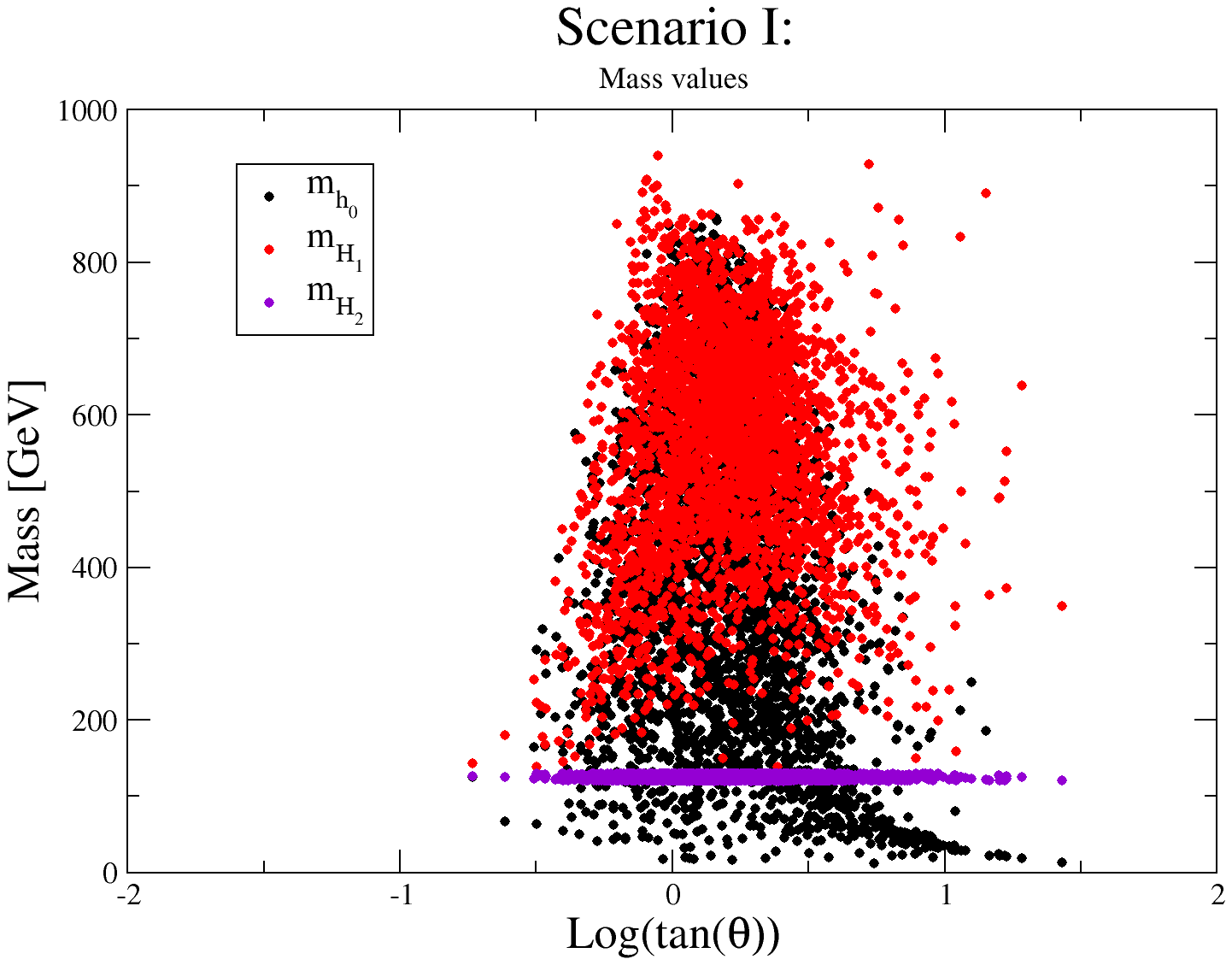}}}
\subfloat[]{{\includegraphics[width=0.48\textwidth]{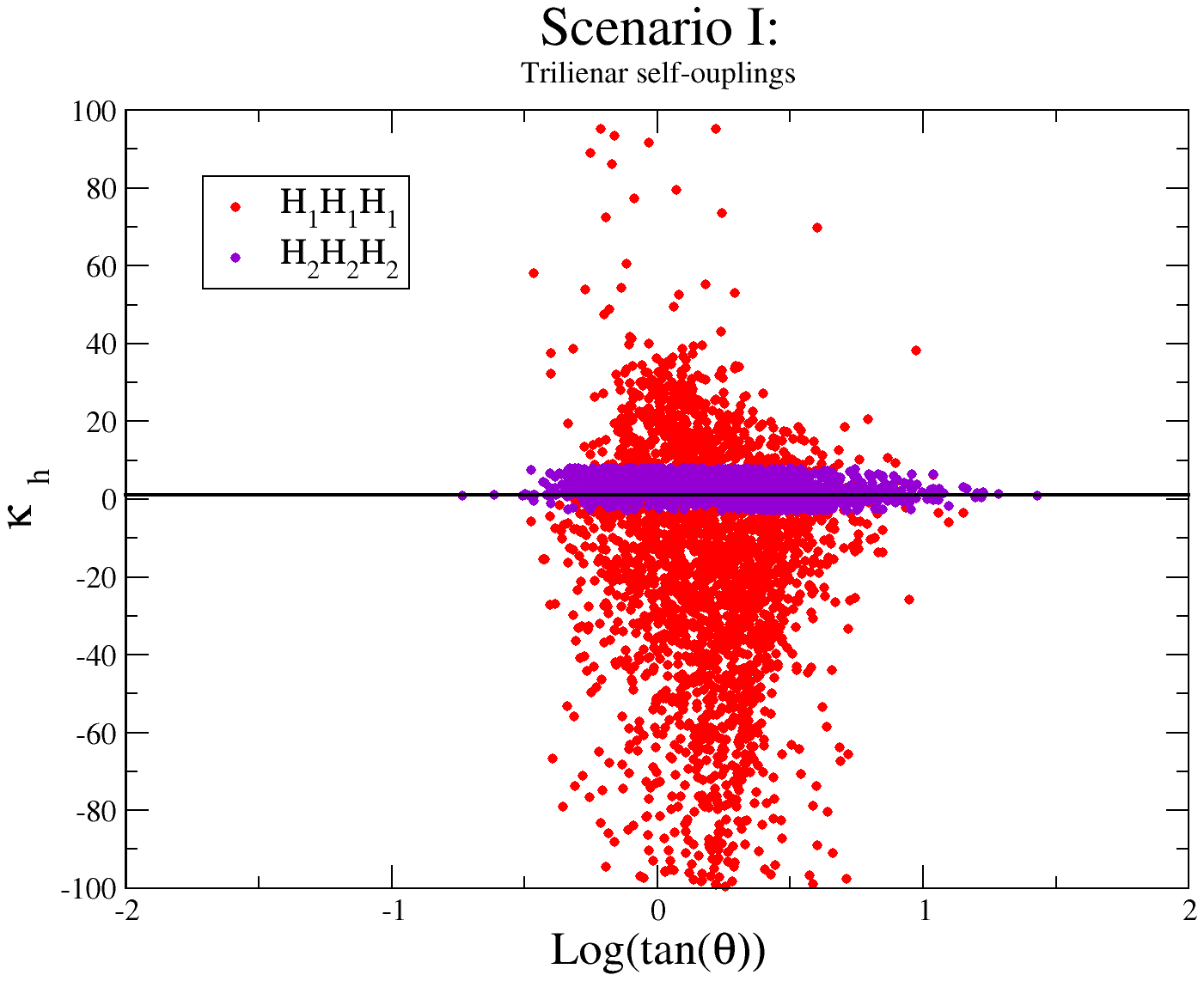}}}\\
\subfloat[]{{\includegraphics[width=0.48\textwidth]{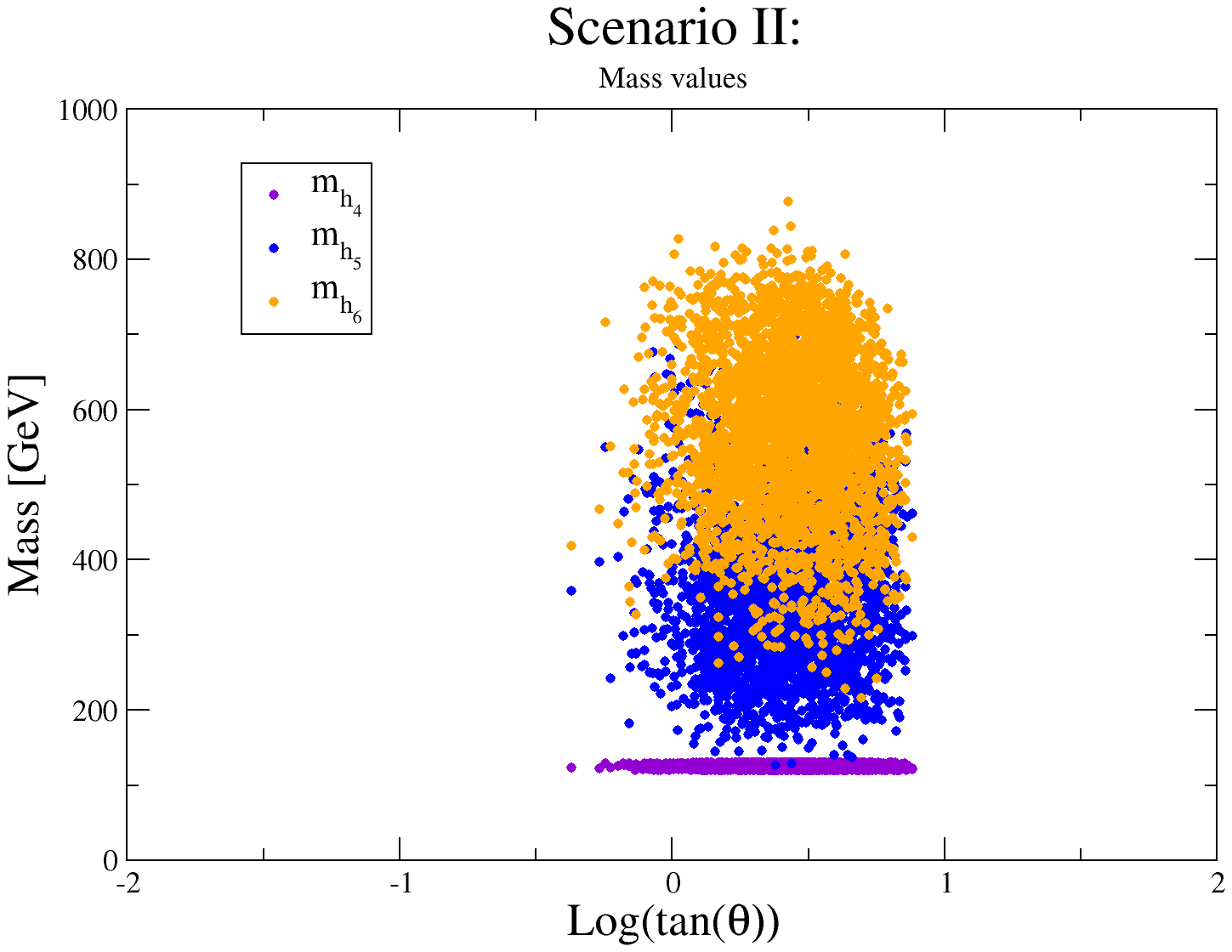}}}
\subfloat[]{{\includegraphics[width=0.48\textwidth]{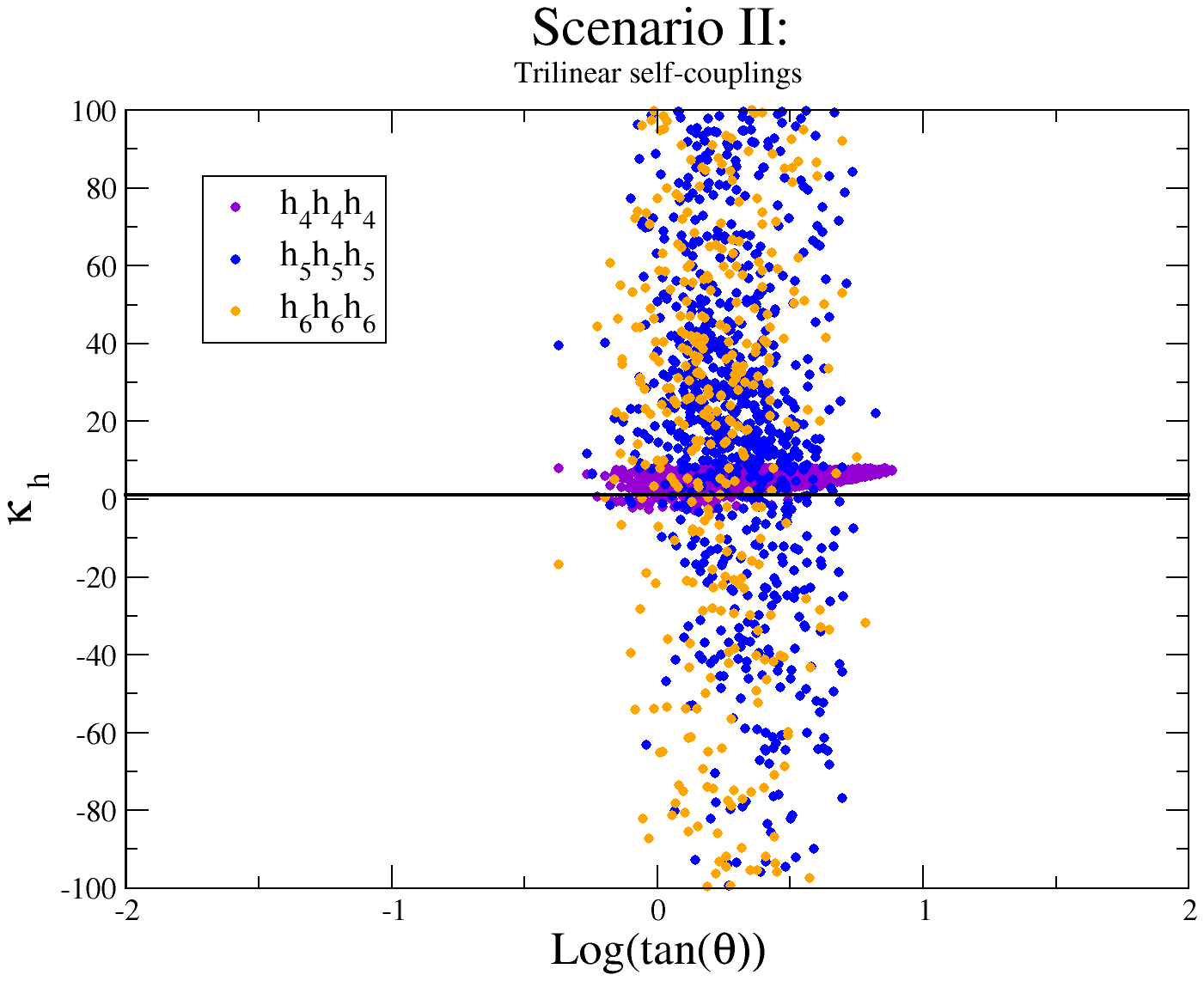}}}\\
\subfloat[]{{\includegraphics[width=0.48\textwidth]{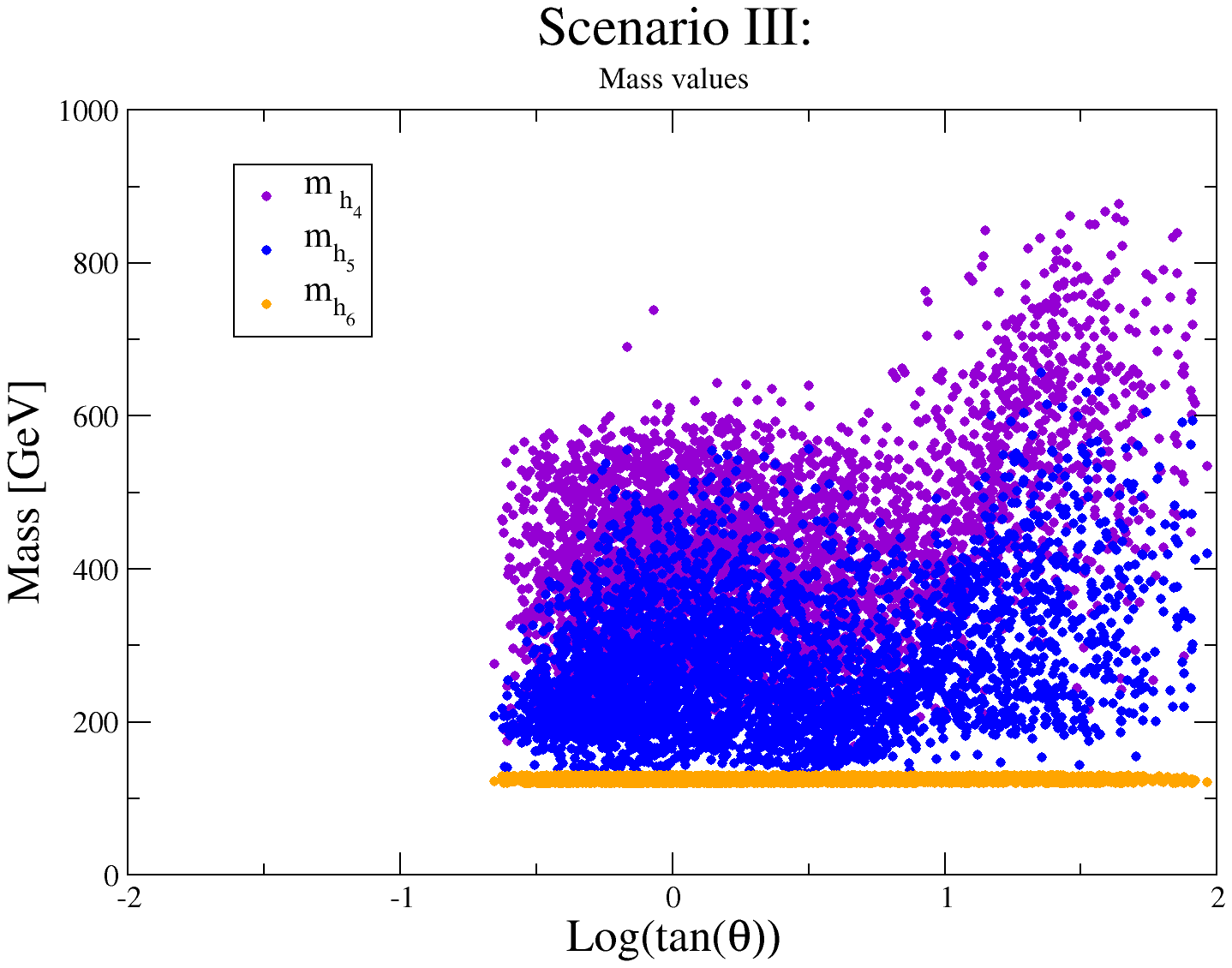}}}
\subfloat[]{{\includegraphics[width=0.48\textwidth]{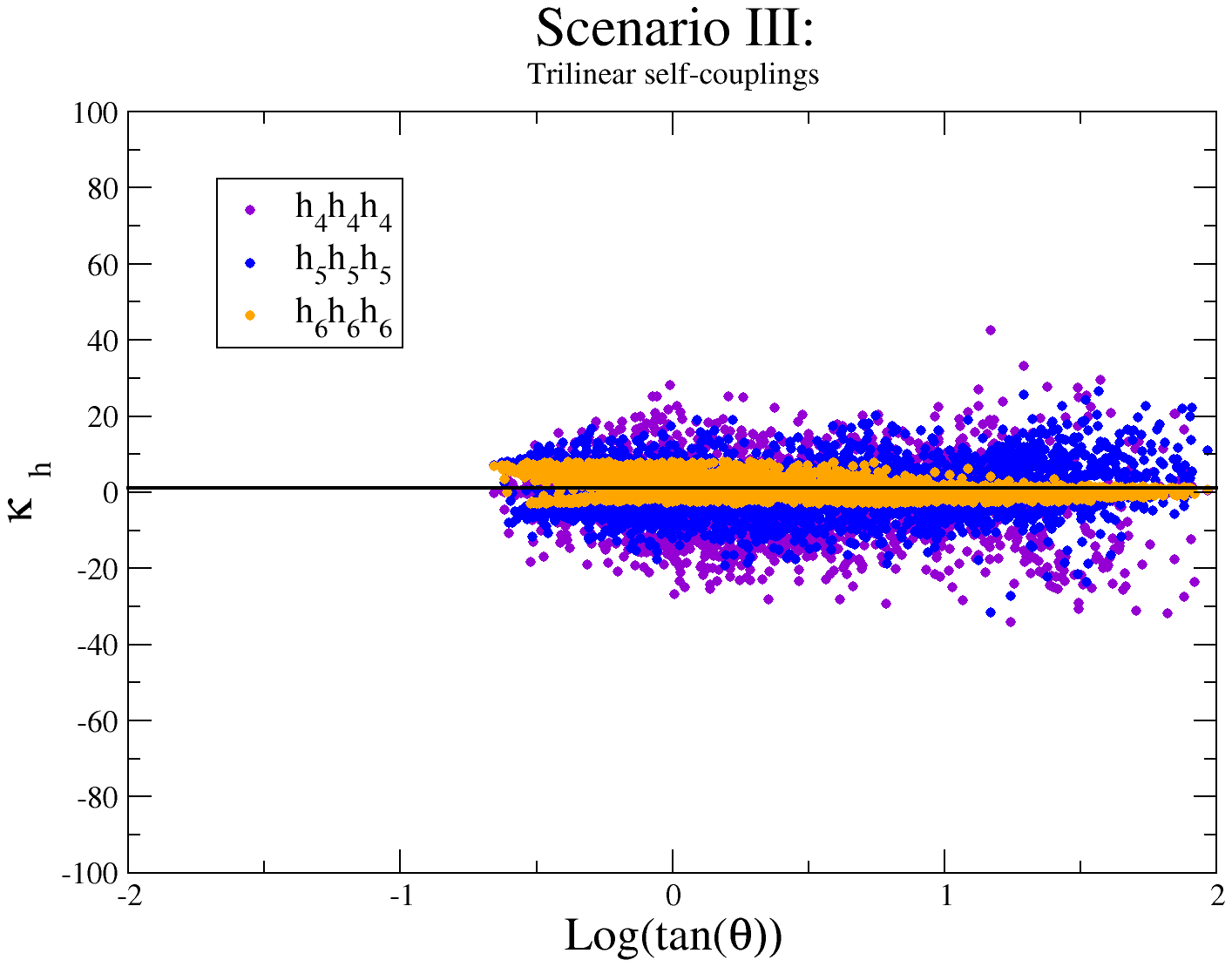}}}
\caption{Plots show Higgs masses spectrum and associated parameter $\kappa_h$ respect to $\textrm{Log}(\text{tan} (\theta))$, for each scenarios: Scenario I, plots (a),(b); Scenario II, plots (c),(d); and Scenario III, plots (e), (f). Black line in plots (a), (c) and (e) is the reference to the values of SM-Higgs mass and black line in plots (b), (d) and (f) is the reference to  1}
\label{fig:S3masstrili}
\end{figure}

For the Scenario II, where we take the complex part of $v_3$  to be nonzero, according to the minimum conditions found, we know that the residual $\mathcal{Z}_2$ symmetry remains, which will cause two of the neutral states to be odd under the symmetry $\mathcal{Z}_2$, which means that the trilinear self-couplings associated to these states are zero, meanwhile the other three states would be nonzero. It is important to remember that in this scenario we have mixed the pseudo-scalar and the scalar states so we can not differentiate from the three states even under the $\mathcal{Z}_2$ symmetry, which ones are pseudo-scalar or scalar states. 
Imposing the restrictions of unitarity, stability, and taking values greater than 80 GeV to the charged Higgs mass, we found that the neutral boson that had more points of its mass around 125 GeV. For this particular scenario, this eigenstate labeled as $h_4$. Therefore, it was to this mass state, which was restricted to take values in the range of 120 to 130 GeV, like as its values for $\kappa_{h}$ between -3 and 8.
 In Figure~\ref{fig:S3masstrili}(c) just $m_{h_{4,5,6}}$ are shown, and the corresponding $\kappa_{h}$ are shown in Figure~\ref{fig:S3masstrili}(d). As we can see, the values for $\textrm{Log}(\text{tan} (\theta))$ are restricted to a range of approximately -0.4 to 0.85, similar behavior to scenario I for the range of values that the angle $\theta$ can be taken.

We can note that the mass range of $m_{h_5}$ and $m_{h_6}$ are values greater than 125 GeV. However $m_{h_6}$ tends to take values greater than 300 GeV, while for their trilinear self-couplings, these states have values very dispersed with respect to the bound that exists for the Higgs of the SM, and as we can note in the range shown, we have fewer points for $h_6$, since these values are above the range shown.

%

\subsection{Scenario III: soft-breaking with explicit CP violation}
In Scenario III, we were able to confirm that adding terms that gently break the $S_3$ symmetry removes the residual symmetry that the model had.  In Figures~\ref{fig:S3masstrili}(e), (f), we just show masses and $\kappa_h$ for $h_{4,5,6}$. Again, we are interested in those that may be candidates for the SM-like Higgs boson, therefore, Higgs states with $m_{h_i}\gtrsim 500\,\textrm{GeV}$ have been removed. By conducting a preliminary analysis and imposing the conditions mentioned in the previous scenarios, we found that, $100 \lesssim m_{h_4} \lesssim 900 \,\textrm{GeV}$, $0\lesssim m_{h_5} \lesssim 600 \,\textrm{GeV}$, and $0 \lesssim m_{h_6} \lesssim 300 \,\textrm{GeV}$. For this reason, we chose the neutral state labeled $h_6$ to restrict its values to the range of 120 to 130 GeV, as well as its values of its trilinear couplings. As we can see in Figure~\ref{fig:S3masstrili}(e)-(f), for this scenario the values that $\textrm{Log}(\text{tan} (\theta))$ takes are between -0.7 and 2, unlike the previous scenarios. The soft breaking terms added widens the range of values that the angle $\theta$ can take, allowing it to reach values close to $\pi/2$. In addition to theses figures we take a particular value to the angle $\varphi= \pi/4$, in order to understand the behavior taking the variation to the parameter $\theta$.     

From Figure~\ref{fig:S3masstrili} (e), we can see that the mass values for states $h_4$ and $h_5$ have a similar range of values. On the other hand, their values for trilinear couplings in this scenario exhibit more restricted behavior compared to the previous scenarios.
In this case, we have $\mathbf{A}$ with the three soft-breaking inputs. The same phase is considered for these one.

\section{Conclusions\label{sec:conclusions}}
We analyzed spontaneous CP breaking and soft-breaking terms provided by the Higgs sector in the SM-$S_3$ framework with an underlying discrete $\mathcal{Z}_2$ symmetry. Three possible scenarios were defined in accordance with the CPV source, which neutral and charged Higgs mass matrices were obtained, and with that the Higgs mass spectrum. Furthermore, trilinear Higgs self-couplings were calculated numerically for the neutral Higgs boson spectrum. In all the three scenarios we found a different allowed space range of the parameters values of the model, which satisfied the conditions imposed like stability and unitarity.  Additionally,  we restricted in each of the scenarios to one of the neutral bosons taking values for its masses in a range of 120 to 130 GeV, as well as its $\kappa_h$ value taking values in a range of -3 to 8, in such a way that this Higgs boson corresponds to the SM-like Higgs boson and to be able to analyze the behavior of the other remanent neutral Higgs bosons.  First, CP-invariant scenario (Scenario I) provided $H_2$ in an allowed parameter region with $\textrm{Log}(\tan(\theta)) \sim 0$  as the SM-like Higgs boson with $m_{H_2}  \ in \sim 125 \, \textrm{GeV}$  and $\kappa_{h} \sim 1$. Second, CPV Scenarios (II and III), with spontaneous CPV given by the $\gamma_3$ phase imposed on the irreducible symmetric representation of the $S_3$ symmetry. 
On the other hand, we explored CPV by inserting three soft-breaking terms into the Higgs potential, provided a spectrum of physical Higgs masses which, $ m_{h_{4}}$ and $ m_{h_{6}}$ respectively, SM-like Higgs boson candidates, with $\kappa_{h} \sim 1$. Therefore, taking into account the corresponding  parameter $\kappa_j$, open up a physical parameter space and give us a possible CPV source in SM-$S_{3}$, in a frame with spontaneous and soft-breaking terms CPV.

\section*{Acknowledgments}

The work of Adriana P\'erez-Mart\'inez is supported by ``Estancias Posdoctorales por M\'exico (SECIHTI)'' and ``Sistema Nacional de Investigadores e Investigadoras'' (SNII-SECIHTI). O.F.B. and E.B.G. acknowledges support from the SECIHTI project No. CBF-2025-G-1187. OF-B, CGH and JH-S thank the support of SNII-SECIHTI, VIEP-BUAP and PRODEP, México.

\appendix

\section{Details of scenario II}\label{apendiceA}
The scenario II corresponds to $\gamma_3 \neq 0$ and $\gamma_1=\gamma_2=0$,
the mass sub-matrices for charged Higgs bosons in Eq.~\eqref{eq:matrizcargada} are given by
\be
{\bf {M_C}}^2_{11}(\gamma_3) = 
\left(
\begin{array}{ccc}
 -4gv_2^2 + \left(4h\cos^2\gamma_3 - \frac{f-h}{2}\right)v_3^2 &
   0 &
   0 \\
 0 & -\frac{1}{2} (f-h)v_3^2 &
   (f-h)\cos\gamma_3 v_2v_3 \\
0 &
(f-h)\cos\gamma_3 v_2v_3 & -2(f-h)v_2^2\\
\end{array}
\right), \label{eq:mc11g3}
\ee
\be
{\bf {M_C}}^2_{12}(\gamma_3) =
\left(
\begin{array}{ccc}
 0 & 0 & 0
   \\
 0 & 0 & -(f-h)\sin\gamma_3 v_2v_3
   \\
0
   &(f-h)\sin\gamma_3 v_2v_3
   & 0 \\
\end{array}
\right). \label{eq:mc21g3}
\ee

Now, we substituted~\eqref{eq:mc11g3} and~\eqref{eq:mc21g3} in~\eqref{eq:matrizcargada}, and diagonalized the resulting matrix. The eigenvalues are 
\be
\left\{0,-\displaystyle\frac{v^2}{2}\left(f-h\right),-\displaystyle\frac{v^2\cos^2\theta}{2}\left(-8h\cos^2\gamma_3 + (f-h)\right)- gv^2\sin^2\theta \right\}.
 \ee
The  neutral Higgs sub-matrices are given by
\be
{\bf {M_N}}^2_{11}(\gamma_3) =
\left(
\begin{array}{ccc}
 9h\cos^2\gamma_3v_3^2 & 0 & 0) \\
0 & 4(c+g)v_2^2 -3h\cos^2\gamma_3 v_3^2  & 2(b+f-2h)\cos\gamma_3 v_2v_3 \\
0 & 2(b+f-2h)\cos\gamma_3 v_2v_3 & 4hv_2^2+a\cos^2\gamma_3 v_3^2 
   \\
\end{array}
\right), \label{eq:mn11g3}
\ee
\be
{\bf {M_N}}^2_{12}(\gamma_3) =
\left(
\begin{array}{ccc}
 3h\cos\gamma_3\sin\gamma_3 v_3^2 & 0 
   & 0 
   \\
 0  & -h\cos\gamma_3\sin\gamma_3 v_3^2  & 2h\sin\gamma_3v_2v_3
   \\
 0 & 2(b+f-h)\sin\gamma_3 v_2v_3
 &  a \cos\gamma_3 \sin\gamma_3 v_3^2 
\end{array}
\right), \label{eq:mn12g3}
\ee
\be
{\bf {M_N}}^2_{22}(\gamma_3) =
\left(
\begin{array}{ccc}
-4(d+g)v_2^2 +h(1+3\cos\gamma_3^2)v_3^2  & 0 & 0 \\
 0 &
   h\sin\gamma_3^2v_3^2 &
   0 \\
 0 & 0 &  a \sin\gamma_3^2v_3^2 \\
\end{array}
\right). \label{eq:mn22g3}
\ee

We computed the neutral matrix~\eqref{eq:matrizneutra}) with~\eqref{eq:mn11g3},~\eqref{eq:mn12g3} and~\eqref{eq:mn22g3}. Diagonalizing the resulting matrix, the eigenvalues are: one zero and five non zero, there are only three Goldstone bosons. When analyzing the  Higgs masses for these three scenarios,  we see again that in scenario 3 the mass spectrum of Higgs bosons is obtained analogous to the normal minimum, where CP is conserved. For this, we have four electrically charged Higgs bosons, with degenerated masses, two by two, five neutral bosons, and three massless bosons, which are given mass to vector bosons. 


\begin{thebibliography}{100}

\bibitem{Higgs:1964pj}
Peter~W. Higgs.
\newblock {Broken Symmetries and the Masses of Gauge Bosons}.
\newblock {\em Phys.Rev.Lett.}, 13:508--509, 1964.

\bibitem{1971NuPhB..35..167T}
G.~{'t Hooft}.
\newblock {Renormalizable Lagrangians for massive Yang-Mills fields}.
\newblock {\em Nuclear Physics B}, 35:167--188, December 1971.

\bibitem{Ishimori:2010au}
Hajime Ishimori, Tatsuo Kobayashi, Hiroshi Ohki, Yusuke Shimizu, Hiroshi Okada,
  et~al.
\newblock {Non-Abelian Discrete Symmetries in Particle Physics}.
\newblock {\em Prog.Theor.Phys.Suppl.}, 183:1--163, 2010.

\bibitem{Ishimori:2012zz}
Hajime Ishimori, Tatsuo Kobayashi, Hiroshi Ohki, Hiroshi Okada, Yusuke Shimizu,
  et~al.
\newblock {An introduction to non-Abelian discrete symmetries for particle
  physicists}.
\newblock {\em Lect.Notes Phys.}, 858:1--227, 2012.

\bibitem{Beye:2015wka}
Florian Beye, Tatsuo Kobayashi, and Shogo Kuwakino.
\newblock Gauge extension of non-abelian discrete flavor.
\newblock {\em JHEP}, 1503:153, 2015.

\bibitem{Derman:1978iz}
Emanuel Derman.
\newblock {Parity Violation in Polarized Electron - Deuteron Scattering Without
  the Parton Model}.
\newblock {\em Phys.Rev.}, D19:133, 1979.

\bibitem{Derman:1979nf}
Emanuel Derman and Hung-Sheng Tsao.
\newblock {SU(2) x U(1) x S(N) flavor dynamics and a bound on the number of
  flavors}.
\newblock {\em Phys. Rev.}, D20:1207, 1979.

\bibitem{Pakvasa:1977in}
S.~Pakvasa and H.~Sugawara.
\newblock Discrete symmetry and cabibbo angle.
\newblock {\em Phys. Lett.}, B73:61, 1978.

\bibitem{Pakvasa:1978tx}
S.~Pakvasa and H.~Sugawara.
\newblock {Mass Of The T Quark In SU(2) X U(1)}.
\newblock {\em Phys. Lett.}, B82:105, 1979.

\bibitem{Mondragon:1998gy}
A.~Mondragon and E.~Rodriguez-Jauregui.
\newblock {The breaking of the flavour permutational symmetry: Mass textures
  and the CKM matrix}.
\newblock {\em Phys. Rev.}, D59:093009, 1999.

\bibitem{Mondragon:1999jt}
A.~Mondragon and E.~Rodriguez-Jauregui.
\newblock {The CP violating phase delta(13) and the quark mixing angles
  Theta(13), Theta(23) and Theta(12) from flavour permutational symmetry
  breaking}.
\newblock {\em Phys. Rev.}, D61:113002, 2000.

\bibitem{Harrison:2003aw}
P.~F. Harrison and W.~G. Scott.
\newblock {Permutation symmetry, tri-bimaximal neutrino mixing and the S3 group
  characters}.
\newblock {\em Phys. Lett.}, B557:76, 2003.

\bibitem{Kubo:2003iw}
J.~Kubo, A.~Mondragon, M.~Mondragon, and E.~Rodriguez-Jauregui.
\newblock {The flavor symmetry}.
\newblock {\em Prog. Theor. Phys.}, 109:795--807, 2003.

\bibitem{Kubo:2003pd}
Jisuke Kubo.
\newblock {Majorana phase in minimal S(3) invariant extension of the standard
  model}.
\newblock {\em Phys. Lett.}, B578:156--164, 2004.

\bibitem{Kobayashi:2003fh}
Tatsuo Kobayashi, Jisuke Kubo, and Haruhiko Terao.
\newblock {Exact S(3) symmetry solving the supersymmetric flavor problem}.
\newblock {\em Phys. Lett.}, B568:83--91, 2003.

\bibitem{Kubo:2004ps}
Jisuke Kubo, Hiroshi Okada, and Fumiaki Sakamaki.
\newblock {Higgs potential in minimal S(3) invariant extension of the standard
  model}.
\newblock {\em Phys. Rev.}, D70:036007, 2004.

\bibitem{Caravaglios:2005gw}
Francesco Caravaglios and Stefano Morisi.
\newblock {Neutrino masses and mixings with an S(3) family permutation
  symmetry}.
\newblock 3 2005.

\bibitem{Araki:2005ec}
Takeshi Araki, Jisuke Kubo, and Emmanuel~A. Paschos.
\newblock {S(3) flavor symmetry and leptogenesis}.
\newblock {\em Eur. Phys. J.}, C45:465--475, 2006.

\bibitem{Kubo:2005sr}
J.~Kubo et~al.
\newblock {A minimal S(3)-invariant extension of the standard model}.
\newblock {\em J. Phys. Conf. Ser.}, 18:380--384, 2005.

\bibitem{Koide:2005ep}
Yoshio Koide.
\newblock {Permutation symmetry S(3) and VEV structure of flavor- triplet Higgs
  scalars}.
\newblock {\em Phys. Rev.}, D73:057901, 2006.

\bibitem{Grimus:2005mu}
Walter Grimus and Luis Lavoura.
\newblock {S(3) x Z(2) model for neutrino mass matrices}.
\newblock {\em JHEP}, 08:013, 2005.

\bibitem{Teshima:2005bk}
T.~Teshima.
\newblock {Flavor mass and mixing and S(3) symmetry: An S(3) invariant model
  reasonable to all}.
\newblock {\em Phys. Rev.}, D73:045019, 2006.

\bibitem{Kimura:2005sx}
T.~Kimura.
\newblock {The minimal S(3) symmetric model}.
\newblock {\em Prog. Theor. Phys.}, 114:329--358, 2005.

\bibitem{Koide:2006vs}
Yoshio Koide.
\newblock {S(3) symmetry and neutrino masses and mixings}.
\newblock {\em Eur. Phys. J.}, C50:809--816, 2007.

\bibitem{Mohapatra:2006pu}
R.~N. Mohapatra, S.~Nasri, and Hai-Bo Yu.
\newblock {S(3) symmetry and tri-bimaximal mixing}.
\newblock {\em Phys. Lett.}, B639:318--321, 2006.

\bibitem{Kaneko:2007ea}
Satoru Kaneko, Hideyuki Sawanaka, Takaya Shingai, Morimitsu Tanimoto, and
  Koichi Yoshioka.
\newblock {New Approach to Texture-zeros with S(3) symmetry - Flavor Symmetry
  and Vacuum Aligned Mass Textures -}.
\newblock 3 2007.

\bibitem{Felix:2006pn}
O.~Felix, A.~Mondragon, M.~Mondragon, and E.~Peinado.
\newblock {Neutrino masses and mixings in a minimal S(3)-invariant extension of
  the standard model}.
\newblock {\em AIP Conf. Proc.}, 917:383--389, 2007.

\bibitem{Canales:2013cga}
F.~Gonz\'alez~Canales, A.~Mondrag\'on, M.~Mondrag\'on, U.~J. Salda\~na Salazar,
  and L.~Velasco-Sevilla.
\newblock {Quark sector of S3 models: classification and comparison with
  experimental data}.
\newblock {\em Phys.Rev.}, D88:096004, 2013.

\bibitem{Emmanuel-Costa:2016vej}
D.~Emmanuel-Costa, O.~M. Ogreid, P.~Osland, and M.~N. Rebelo.
\newblock {Spontaneous symmetry breaking in the $S_3$-symmetric scalar sector}.
\newblock {\em JHEP}, 02:154, 2016.
\newblock [Erratum: JHEP 08, 169 (2016)].

\bibitem{Vien:2014vka}
V.~V. Vien and H.~N. Long.
\newblock {Neutrino mass and mixing in the 3-3-1 model and $S_3$ flavor
  symmetry with minimal Higgs content}.
\newblock {\em Zh. Eksp. Teor. Fiz.}, 145:991--1009, 2014.

\bibitem{Mondragon:2007af}
A.~Mondragon, M.~Mondragon, and E.~Peinado.
\newblock {Lepton masses, mixings and FCNC in a minimal $S_3$-invariant
  extension of the Standard Model}.
\newblock {\em Phys. Rev.}, D76:076003, 2007.

\bibitem{Beltran:2009zz}
O.~Felix Beltran, M.~Mondragon, and E.~Rodriguez-Jauregui.
\newblock {Conditions for vacuum stability in an S(3) extension of the standard
  model}.
\newblock {\em J. Phys. Conf. Ser.}, 171:012028, 2009.

\bibitem{Barger:2009me}
Vernon Barger, Heather~E. Logan, and Gabe Shaughnessy.
\newblock {Identifying extended Higgs models at the LHC}.
\newblock {\em Phys. Rev.}, D79:115018, 2009.

\bibitem{Gupta:2009wn}
Rick~S. Gupta and James~D. Wells.
\newblock {Next Generation Higgs Bosons: Theory, Constraints and Discovery
  Prospects at the Large Hadron Collider}.
\newblock {\em Phys.Rev.}, D81:055012, 2010.

\bibitem{Kanemura:2004mg}
Shinya Kanemura, Yasuhiro Okada, Eibun Senaha, and C.~P. Yuan.
\newblock {Higgs coupling constants as a probe of new physics}.
\newblock {\em Phys. Rev.}, D70:115002, 2004.

\bibitem{Djouadi:2005gj}
Abdelhak Djouadi.
\newblock {The Anatomy of electro-weak symmetry breaking. II. The Higgs bosons
  in the minimal supersymmetric model}.
\newblock {\em Phys. Rept.}, 459:1--241, 2008.

\bibitem{Branco:2011iw}
G.C. Branco, P.M. Ferreira, L.~Lavoura, M.N. Rebelo, Marc Sher, et~al.
\newblock {Theory and phenomenology of two-Higgs-doublet models}.
\newblock {\em Phys.Rept.}, 516:1--102, 2012.

\bibitem{Lendvai:1981wn}
E.~Lendvai and G.~Pocsik.
\newblock {UPPER BOUNDS ON HIGGS BOSON MASSES IN THE WEINBERG-SALAM MODEL WITH
  THREE HIGGS DOUBLETS}.
\newblock {\em Phys. Lett.}, B106:314, 1981.

\bibitem{Adler:1999gv}
Stephen~L. Adler.
\newblock {Higgs mass bounds in the three- and six-Higgs doublet models for
  family structure}.
\newblock {\em Phys. Rev.}, D60:015002, 1999.

\bibitem{Ferreira:2008zy}
P.~M. Ferreira and Joao~P. Silva.
\newblock {Discrete and continuous symmetries in multi-Higgs-doublet models}.
\newblock {\em Phys. Rev.}, D78:116007, 2008.

\bibitem{Howl:2009ds}
R.~Howl and S.F. King.
\newblock {Solving the Flavour Problem in Supersymmetric Standard Models with
  Three Higgs Families}.
\newblock {\em Phys.Lett.}, B687:355--362, 2010.

\bibitem{Barroso:2005tq}
A.~Barroso, P.~M. Ferreira, and R.~Santos.
\newblock {Some remarks on tree-level vacuum stability in two Higgs doublet
  models}.
\newblock {\em Afr. J. Math. Phys.}, 3:103--109, 2006.

\bibitem{Barroso:2005da}
A.~Barroso, P.~M. Ferreira, and R.~Santos.
\newblock {Tree-level vacuum stability in multi Higgs models}.
\newblock {\em PoS}, HEP2005:337, 2006.

\bibitem{Barroso:2006pa}
A.~Barroso, P.~M. Ferreira, R.~Santos, and Joao~P. Silva.
\newblock {Stability of the normal vacuum in multi-Higgs-doublet models}.
\newblock {\em Phys. Rev.}, D74:085016, 2006.

\bibitem{Hagedorn:2006ir}
C.~Hagedorn, M.~Lindner, and F.~Plentinger.
\newblock {The discrete flavor symmetry D(5)}.
\newblock {\em Phys. Rev.}, D74:025007, 2006.

\bibitem{Tofighi:2009zzb}
A.~Tofighi and M.~Moazzen.
\newblock {Neutral minima in three-Higgs doublet models}.
\newblock {\em Int. J. Theor. Phys.}, 48:3372--3382, 2009.

\bibitem{Morisi:2009sc}
S.~Morisi and E.~Peinado.
\newblock {An A4 model for lepton masses and mixings}.
\newblock {\em Phys. Rev.}, D80:113011, 2009.

\bibitem{Morisi:2010rk}
S.~Morisi and E.~Peinado.
\newblock {An S4 model for quarks and leptons with maximal atmospheric angle}.
\newblock {\em Phys. Rev. D}, 81:085015, 2010.

\bibitem{EmmanuelCosta:2007zz}
D.~Emmanuel-Costa, O.~Felix-Beltran, M.~Mondragon, and E.~Rodriguez-Jauregui.
\newblock {Stability of the tree-level vacuum in a minimal S(3) extension of
  the standard model}.
\newblock {\em AIP Conf. Proc.}, 917:390--393, 2007.

\bibitem{Dev:2012ns}
S.~Dev, Radha~Raman Gautam, and Lal Singh.
\newblock {Broken $S_3$ Symmetry in the Neutrino Mass Matrix and Non-Zero
  $\theta_{13}$}.
\newblock {\em Phys.Lett.}, B708:284--289, 2012.

\bibitem{Dias:2012bh}
A.G. Dias, A.C.B. Machado, and C.C. Nishi.
\newblock {An $S_3$ Model for Lepton Mass Matrices with Nearly Minimal
  Texture}.
\newblock {\em Phys.Rev.}, D86:093005, 2012.

\bibitem{Canales:2012dr}
F.~Gonzalez~Canales, A.~Mondragon, and M.~Mondragon.
\newblock {The $S_3$ Flavour Symmetry: Neutrino Masses and Mixings}.
\newblock {\em Fortsch.Phys.}, 61:546--570, 2013.

\bibitem{Aad:2013xqa}
Georges Aad et~al.
\newblock {Evidence for the spin-0 nature of the Higgs boson using ATLAS data}.
\newblock {\em Phys.Lett.}, B726:120--144, 2013.

\bibitem{Aad:2012tfa}
Georges Aad et~al.
\newblock {Observation of a new particle in the search for the Standard Model
  Higgs boson with the ATLAS detector at the LHC}.
\newblock {\em Phys.Lett.}, B716:1--29, 2012.

\bibitem{Chatrchyan:2013lba}
Serguei Chatrchyan et~al.
\newblock {Observation of a new boson with mass near 125 GeV in pp collisions
  at $\sqrt{s}$ = 7 and 8 TeV}.
\newblock {\em JHEP}, 1306:081, 2013.

\bibitem{Chatrchyan:2012ufa}
Serguei Chatrchyan et~al.
\newblock {Observation of a new boson at a mass of 125 GeV with the CMS
  experiment at the LHC}.
\newblock {\em Phys.Lett.}, B716:30--61, 2012.

\bibitem{Miller:2000uz}
D.J. Miller.
\newblock {Can the triple Higgs selfcoupling be measured at future colliders?}
\newblock {\em Nucl.Phys.Proc.Suppl.}, 89:70--75, 2000.

\bibitem{Baur:2003gp}
U.~Baur, T.~Plehn, and David~L. Rainwater.
\newblock {Probing the Higgs selfcoupling at hadron colliders using rare
  decays}.
\newblock {\em Phys.Rev.}, D69:053004, 2004.

\bibitem{Dutta:2008bh}
Sukanta Dutta, Kaoru Hagiwara, and Yu~Matsumoto.
\newblock {Measuring the Higgs-Vector boson Couplings at Linear $e^{+} e^{-}$
  Collider}.
\newblock {\em Phys.Rev.}, D78:115016, 2008.

\bibitem{Barr:2014sga}
Alan~J. Barr, Matthew~J. Dolan, Christoph Englert, Danilo~Enoque Ferreira~de
  Lima, and Michael Spannowsky.
\newblock {Higgs Self-Coupling Measurements at a 100 TeV Hadron Collider}.
\newblock {\em JHEP}, 1502:016, 2015.

\bibitem{Cheng:2025aev}
Alkaid Cheng.
\newblock {Combination of searches for resonant Higgs boson pair production
  using $pp$ collisions at $\sqrt s$ = 13 TeV with the ATLAS detector}.
\newblock {\em PoS}, LHCP2024:248, 2025.

\bibitem{ATLAS:2023gzn}
Georges Aad et~al.
\newblock {Studies of new Higgs boson interactions through nonresonant HH
  production in the $ b\overline{b}\gamma \gamma $ final state in pp collisions
  at $ \sqrt{s} $ = 13 TeV with the ATLAS detector}.
\newblock {\em JHEP}, 01:066, 2024.

\bibitem{Hayrapetyan_2025}
Aram Hayrapetyan et~al.
\newblock {Constraints on the Higgs boson self-coupling from the combination of
  single and double Higgs boson production in proton-proton collisions at
  s=13TeV}.
\newblock {\em Phys. Lett. B}, 861:139210, 2025.

\bibitem{2025139210}
A.~Hayrapetyan et~al.
\newblock Constraints on the higgs boson self-coupling from the combination of
  single and double higgs boson production in proton-proton collisions at
  s=13tev.
\newblock {\em Physics Letters B}, 861:139210, 2025.

\bibitem{Barradas-Guevara:2014yoa}
E.~Barradas-Guevara, O.~Felix-Beltran, and E.~Rodriguez-Jauregui.
\newblock {Trilinear self-couplings in an S(3) flavored Higgs model}.
\newblock {\em Phys.Rev.}, D90(9):095001, 2014.

\bibitem{Bhattacharyya:2010hp}
Gautam Bhattacharyya, Philipp Leser, and Heinrich Pas.
\newblock {Exotic Higgs boson decay modes as a harbinger of $S_3$ flavor
  symmetry}.
\newblock {\em Phys.Rev.}, D83:011701, 2011.

\bibitem{Bhattacharyya:2012ze}
Gautam Bhattacharyya, Philipp Leser, and Heinrich Pas.
\newblock {Novel signatures of the Higgs sector from S3 flavor symmetry}.
\newblock {\em Phys.Rev.}, D86:036009, 2012.

\bibitem{Das:2014fea}
Dipankar Das and Ujjal~Kumar Dey.
\newblock {Analysis of an extended scalar sector with $S_3$ symmetry}.
\newblock {\em Phys. Rev. D}, 89(9):095025, 2014.
\newblock [Erratum: Phys.Rev.D 91, 039905 (2015)].

\bibitem{Gomez-Bock:2021uyu}
M.~G{\'o}mez-Bock, M.~Mondrag{\'o}n, and A.~P{\'e}rez-Mart{\'\i}nez.
\newblock {Scalar and gauge sectors in the 3-Higgs Doublet Model under the
  $S_3$ symmetry}.
\newblock {\em Eur. Phys. J. C}, 81(10):942, 2021.

\bibitem{Lee:1974jb}
T.D. Lee.
\newblock {CP Nonconservation and Spontaneous Symmetry Breaking}.
\newblock {\em Phys.Rept.}, 9:143--177, 1974.

\bibitem{Chen:2015gaa}
Chien-Yi Chen, S.~Dawson, and Yue Zhang.
\newblock {Complementarity of LHC and EDMs for Exploring Higgs CP Violation}.
\newblock {\em JHEP}, 06:056, 2015.

\bibitem{VALIAHO198619}
Hannu V{\"a}liaho.
\newblock Criteria for copositive matrices.
\newblock {\em Linear Algebra and its Applications}, 81:19 -- 34, 1986.

\bibitem{Unwin:2011rn}
James Unwin.
\newblock {Vacuum stability and the Cholesky decomposition}.
\newblock {\em Eur. Phys. J.}, C71:1663, 2011.

\bibitem{Gunion:2005ja}
John~F. Gunion and Howard~E. Haber.
\newblock {Conditions for CP-violation in the general two-Higgs-doublet model}.
\newblock {\em Phys. Rev. D}, 72:095002, 2005.

\bibitem{Kuncinas:2023ycz}
A.~Kun{\v{c}}inas, O.~M. Ogreid, P.~Osland, and M.~N. Rebelo.
\newblock {Complex S$_{3}$-symmetric 3HDM}.
\newblock {\em JHEP}, 07:013, 2023.

\bibitem{CMS:2024awa}
Aram Hayrapetyan et~al.
\newblock {Constraints on the Higgs boson self-coupling from the combination of
  single and double Higgs boson production in proton-proton collisions at
  s=13TeV}.
\newblock {\em Phys. Lett. B}, 861:139210, 2025.

\end{thebibliography}

\end{document}